\documentclass[10pt,epsf]{article}

\usepackage{graphicx}
\usepackage{indentfirst}

\begin{document}

\title{\bf{Rotating Black Hole Thermodynamics with a Particle Probe}}
\date{}
\maketitle

\begin{center}\author{Bogeun Gwak}\footnote{rasenis@sogang.ac.kr} and \author{Bum-Hoon Lee}\footnote{bhl@sogang.ac.kr}\\ \vskip 0.25in $^{1,2}$\it{Department of Physics and Center of Quantum Spacetime, Sogang University, Seoul 121-741, Korea} \end{center} \vskip 0.6in

{\abstract
{The thermodynamics of Myers-Perry black holes in general dimensions are studied using a particle probe. When undergoing particle absorption, the changes of the entropy and irreducible mass are shown to be dependent on the particle radial momentum. The black hole thermodynamic behaviors are dependent on dimensionality for specific rotations. For a 4-dimensional Kerr black hole, its black hole properties are maintained for any particle absorption. 5-dimensional black holes can avoid a naked ring singularity by absorbing a particle in specific momenta ranges. Black holes over 6 dimensions become ultra-spinning black holes through a specific form of particle absorption. The microscopical changes are interpreted in limited cases of Myers-Perry black holes using Kerr/CFT correspondence. We systematically describe the black hole properties changed by particle absorption in all dimensions.}}

\thispagestyle{empty}
\newpage

\setcounter{page}{1}
\section{Introduction}
Black hole thermodynamic properties are affected by matter absorption. Using the integrable structure of geodesics in the 4-dimensional Kerr black hole\,\cite{Carter}, the relation between a particle absorbed and Kerr black hole charges is obtained. The absorption increases the irreducible mass\,\cite{ChrisRu} of the black hole, which is interpreted as its surface energy\,\cite{Smarr}. Since the irreducible mass has a one-to-one correspondence with the black hole area, the particle absorption process relates to the entropy. The experiments of the Large Hadron Collider (LHC) or future colliders may give a clue as to the conceivable possible extra dimensions. To see the roles of these extra dimensions in the black hole, generalization of the black hole properties to higher-dimensions is needed. The additional spatial freedom gives a variety of black hole horizon topologies and extends the expected types of stable black holes\,\cite{vbh}. This makes it possible to find non-spherical horizons and multi-black holes. In higher dimensions, a class of higher dimensional rotating black holes can be described in terms of a phase diagram in entropy and spin. Compared with black rings and black Saturns, the Myers-Perry (MP) black hole\,\cite{MP} has the highest entropy at lower spin, and the smallest at higher spin\,\cite{emp1,emp2}. As a higher dimensional generalization of the Kerr black hole, the MP black holes can extend the four dimensional black hole thermodynamics to that of a higher dimensional one. Some effects of particle absorption in the MP black holes has been studied\,\cite{BHcen}, but the thermodynamic properties and phase structure due to particle absorption are not yet elucidated. Through the integrable structure of particle geodesics in MP black holes\,\cite{VPage1,VPage2}, the change of MP black hole properties due to the absorption process can be obtained as a generalized extension of the Kerr black hole. In this work, we will investigate the particle absorption, relate it to the thermodynamic processes in the higher-dimensional MP black holes, and formulate the process as a unified expression in all dimensions. The changes of the irreducible mass and entropy are constructed in terms of the particle momenta. Black holes interacts to some fields in the universe\,\cite{astro}. When MP black holes absorb a specific particle, our possible observations about the changes of MP black hole properties and how much these are dependent on the spacetime dimensions will be shown. From the microscopic point of view, the absorbing particle energy contributes to the energy degeneracy of the black hole and increases the Bekenstein-Hawking entropy\,\cite{BH1}. On the string theory side, the entropy can be explained by the logarithm of the micro-state counting of the D-brane configurations in the specific black hole\,\cite{entD}. As an extension of AdS/CFT correspondence\,\cite{acc1}, the Bekenstein-Hawking entropy of the Kerr black hole is interpreted as a microscopic entropy for the specific conformal field theory(CFT), which leads to the conjecture of Kerr/CFT correspondence\,\cite{kerrCFT}. In this context, the conversion of the particle momenta to that of the MP black holes can reach the micro-state of the MP black hole and opens up the possibility of higher dimensional generalization of the Kerr/CFT correspondence.

In this paper, we probe the black hole thermodynamic properties by a particle absorption. In the most of cases, the black hole properties are compared along to one variable change. However, if the mass and angular momenta are changed by hand on the metric, these can cause to decrease the entropy or change other properties as unphysical ways. Applied this point of view to the black hole, comparing with between different black hole masses or spins means watching different black holes in the entropy point of view, because the increase of spin parameter with fixed black hole mass causes the decrease of irreducible mass, and also similar situation happened in the changing black hole mass. To avoid and interpret relevantly these problems, we apply a particle absorption which the real black hole undergoes. The black hole mass and momenta can be controlled by the absorption of a particle carrying specific momenta. Using this procedure, the black hole properties are changed in continuous way. This method can change black hole properties in the satisfaction of thermodynamic 2nd law. Also it is what the astrophysical black holes undergo in the universe. The particle changes the black hole mass and momenta at the same time. Conceptually, it is the particle perturbation to the black hole. The black hole mass, angular momenta, and horizon size are organically connected to each other and will be show to be how much these are changed after particle absorption as the thermodynamic changes of the Myers-Perry black hole. Our analysis coincidences to what observers can do for the black hole in the real world. For astrophysical black hole, since observations are localized in 4-dimensional spacetime, we can obtain clues about extra dimensions through expected behaviors of black holes by falling the particles as a probe. Also, Myers-Perry black holes include a simple rotating black hole like Kerr black holes or black membranes. Therefore, we can describe the thermodynamic property changes can be observed in these all black holes.

The paper is organized as follows. First, the thermodynamic process of the particle absorption is considered in $N$-dimensional MP black holes with a single rotating plane which also include Kerr black hole in $N$=4\,. Secondly, the particle probes the thermodynamic transformations in the $N$-dimensional MP black holes with arbitrary rotations. Thirdly, the thermodynamic changes of the MP black holes are explained systematically from the point of view of particle absorption. Fourthly, we show the evolution of MP black holes in specific cases. Fifthly, the effects of the particle absorption to microstates is interpreted in the Kerr/CFT correspondence point of view. Sixthly, we summarize and discuss our results.

\section{The Particle Absorption in Black holes with a Single Rotation}
To extend the thermodynamic process to the $N$-dimensional cases, we consider particle absorption in the $N$-dimensional MP black holes with a single rotation. The role of additional spatial dimensions can be proved in terms of the particle absorption. The spacetime consists of 4 dimensional black hole with an additional ($N$-4)-dimensional sphere. Note that the MP black holes become Kerr black holes when $N$=4\,, so all results in this section can be applied to Kerr black holes, too. The metric is\,\cite{MP,emp1},
\begin{eqnarray}
ds^2=-dt^2+\frac{2M}{r^{N-5}\Sigma}(dt-a\sin^2\theta d\phi)^2+\frac{\Sigma}{\Delta}dr^2+\Sigma d\theta^2+(r^2+a^2)\sin^2\theta d\phi^2+r^2\cos^2\theta\sum_{i=1}^{N-4}\prod_{j=1}^{i-1}\sin^2\psi_j d\psi_i^2\,.
\end{eqnarray}
where $\Sigma=r^2+a^2\cos^2\theta\,$ and $\Delta=r^2+a^2-\frac{2M}{r^{N-5}}$. Note that the coordinates $\psi_i$s represent a $(N-4)$-dimensional sphere, $\Omega_{N-4}$\,. Generally, the orbits of the massive or massless particles are described by geodesic equations\,\cite{Gwak}. In this work, it is needed to construct first-order geodesic equations in order to determine the particle energy through the particle momenta. We apply the Hamilton-Jacobi method\,\cite{Carter} to the MP black holes, then the first order geodesics are obtained using the integrability. The Hamiltonian is written, 
\begin{eqnarray}
\mathcal{H}=\frac{1}{2}g^{\alpha\beta}p_\alpha p_\beta\,.
\end{eqnarray}
The metric has translation symmetries for $t$, $\phi$, and $\psi_{N-4}$\,, and the corresponding conserved charges are defined as $E$, $L$, and $R_{N-4}$. The Hamilton-Jacobi action is constructed as,
\begin{eqnarray}
S=\frac{1}{2}m^2\lambda-Et+L\phi+S_r(r)+S_\theta(\theta)+\sum_{i=1}^{N-5}S_{\psi_i}(\psi_i)+R_{N-4}\psi_{N-4}\,,
\end{eqnarray}
in which $m$ is particle mass. The inverse metric of $N$-dimensional MP black hole is given,
\begin{eqnarray}
&&g^{tt}=-1-\frac{2(a^2+r^2)M}{\Delta\,\Sigma\,r^{N-5}}\,,\,\,g^{t\phi}=g^{\phi t}=-\frac{2aM}{\Delta\,\Sigma\,r^{N-5}}\,,\,\,g^{\phi\phi}=\frac{\csc^2\theta}{\Sigma}-\frac{a^2}{\Delta\,\Sigma}\,,\\
&&g^{rr}=\frac{\Delta}{\Sigma}\,,\,\,g^{\theta\theta}=\frac{1}{\Sigma}\,,\,\,g^{\psi_i\psi_i}=\frac{\sec^2\theta}{r^2}\prod_{j=1}^{i-1}\csc^2\psi_j\,.\nonumber
\end{eqnarray}
By integrability, the Hamilton-Jacobi equations are separated,
\begin{eqnarray}
&&\mathcal{K}=(m^2-E^2)r^2-\frac{2(r^2+a^2)E^2 M}{\Delta r^{N-5}}+\frac{4aELM}{\Delta r^{N-5}}-\frac{a^2 L^2}{\Delta}+\Delta \left(\frac{dS_r}{dr}\right)^2+\frac{a^2}{r^2}R_1^2\,,\\
&&\mathcal{K}=-(m^2-E^2)a^2\cos^2\theta -L^2\csc^2\theta-\left(\frac{dS_\theta}{d\theta}\right)^2-\sec^2\theta R_1^2\,,\,\,\,\,R_i=\left(\frac{dS_{\psi_i}}{d\psi_i}\right)^2+R_{i+1}^2\csc^2\psi_i\,,\nonumber
\end{eqnarray}
where the separate constants are $\mathcal{K}$ and $R_i$s. The rewritten action is,
\begin{eqnarray}
\label{eq:nsingle1}
&&S=\frac{1}{2}m^2\lambda-Et+L\phi+\int\sqrt{R(r)}dr+\int\sqrt{\Theta(\theta)}d\theta+\sum_{i=1}^{N-5}\int\sqrt{\Psi_i(\psi_i)}d\psi_i+R_{N-4}\psi_{N-4}\,,\\
&&R(r)=-\frac{1}{\Delta}\left[(m^2-E^2)r^2+\frac{a^2}{r^2}R_1^2\right]+\frac{2(r^2+a^2)E^2M}{\Delta^2 r^{N-5}}-\frac{4aELM}{\Delta^2 r^{N-5}}+\frac{a^2 L^2}{\Delta^2}+\frac{\mathcal{K}}{\Delta}\,,\nonumber\\
&&\Theta(\theta)=-(m^2-E^2)a^2\cos^2\theta -L^2\csc^2\theta-\sec^2\theta R_1^2-\mathcal{K}\,,\,\,\,\,\Psi_i=R_i^2-R_{i+1}^2\csc^2\psi_i\,.\nonumber
\end{eqnarray}
The full of geodesics can be obtained from eq.~(\ref{eq:nsingle1}). The particle energy equation in specific locations and momenta is constructed using geodesics of $r$ and $\theta$ coordinates,
\begin{eqnarray}
p^{r}=\dot{r}=\frac{\Delta}{\Sigma}\sqrt{R}\,,\,\,\,\,p^{\theta}=\dot{\theta}=\frac{1}{\Sigma}\sqrt{\Theta}\,.
\end{eqnarray}
The particle energy is determined by,
\begin{eqnarray}
&&\alpha E^2 +\beta E + \gamma=0\,,\\
&&\alpha=\Sigma+\frac{2(r^2+a^2)M}{\Delta r^{N-5}}\,,\,\,\,\,\beta=-\frac{4aLM}{\Delta r^{N-5}}\,,\nonumber\\
&&\gamma=-m^2\Sigma-\frac{a^2}{r^2}R_1^2+a^2L^2-L^2\csc^2\theta-R_1^2\sec^2\theta-\Sigma^2\left(\frac{(p^r)^2}{\Delta}+(p^\theta)^2\right)\,.\nonumber
\end{eqnarray}
The constant $\alpha$ is always positive outside the horizon, $r_h$. A positive-root solution is selected as representing the future-forwarding particle energy\,\cite{ChrisRu}\,. For a given $r>r_h$, the massive particle energy is bigger than that of the massless particle by a positive contribution of the mass term. The particle energy is applied to the black hole when the particle reaches the black hole outer horizon. The particle radial and angular momenta are important in determining the particle energy at the horizon. The particle energy is obtained,
\begin{eqnarray}
\label{eq:NminE}
E=\frac{aL}{r_h^2+a^2}+\frac{|\Sigma p^r|}{r_h^2+a^2}\geq E_{min}=\frac{aL}{r_h^2+a^2}\,.
\end{eqnarray}
where the minimum energy is achieved when $p^r=0$. Note that the particle energy is dependent on the coordinate $\theta\,$ in $\Sigma$\,. The black hole mass $M_B$ and angular momentum $J$ are changed by the particle energy and the angular momentum, then $\delta M_B\,=\,E\,$ and $\delta J\,=\,L\,$. Additional spacetime information is implicitly included in $r_h$. Additional momenta, $R_i$, cannot change that of the black hole. From eq.~(\ref{eq:NminE}), the black hole mass change is rewritten explicitly,
\begin{equation}
\label{eq:delta_M}
\delta M_B=\frac{a\delta J}{r_h^2+a^2}+\frac{|\Sigma p^r|}{r_h^2+a^2}\geq E_{min},
\end{equation}
where $M_B=\frac{(N-2)\Omega_{N-2}}{8\pi G}M$ and $J=\frac{\Omega_{N-2}}{4\pi G}Ma$ are the black hole mass and angular momentum. Note that the constant $G$ is 1 for simplicity. The black hole mass change is dependent on the sum of $|\Sigma p^r|$ and $aL$\,. The change of the irreducible mass $M_{ir}$ is obtained by removing the black hole rotation energy contribution by the particle. The irreducible mass and its change are obtained,
\begin{eqnarray}
\label{eq:delta_Mir}
M_{ir}=\sqrt{r_h^{N-4}(r_h^2+a^2)}=\sqrt{2Mr_h}\,,\,\,\,\,\delta M_{ir}=\frac{8\pi G r_h |\Sigma p^r |(r_h^{N-4}(r_h^2+a^2))^{-\frac{1}{2}}}{\Omega_{N-2}((N-3)r_h^2+(N-5)a^2)}\,,
\end{eqnarray}
where $\delta M_{ir}$ is always positive for $N\geq 5$ and for the $N=4$ Kerr case in $M \geq a$\,. Using eq.~(\ref{eq:delta_M}) and (\ref{eq:delta_Mir}), the horizon and spin parameter changes are obtained as particle momenta,
\begin{eqnarray}
\label{deltaarh1}
&&\delta a=\frac{1}{M_B}\left[-\frac{a\Sigma}{r_h^2+a^2}|p^r|+\frac{(N-2)r_h^2+(N-4)a^2}{2(r_h^2+a^2)}L\right]\,,\\ 
&&\delta r_h = \frac{1}{M_B}\left[\frac{(r_h^2+3a^2)r_h\Sigma}{(r_h^2+a^2)((N-3)r_h^2+(N-5)a^2)}|p^r|-\frac{a r_h}{r_h^2 + a^2}L\right]\,.\nonumber
\end{eqnarray}
The area of the black hole is proportional to the square of the irreducible mass as,
\begin{eqnarray}
\mathcal{A}=\Omega_{N-2}r_h^{N-4}(r_h^2+a^2)=\Omega_{N-2}M_{ir}^2\,.
\end{eqnarray}
The Bekenstein-Hawking entropy is written in terms of irreducible mass. The entropy change in the absorption is obtained using the one-to-one correspondence, 
\begin{eqnarray}
\label{singleS1}
S_{BH}=\frac{1}{4}\Omega_{N-2}M_{ir}^2\,,\,\,\delta S_{BH}=\frac{4\pi r_h |\Sigma p^r |}{(N-3)r_h^2+(N-5)a^2}\,,
\end{eqnarray}
where the Kerr black hole can be obtained when $N=4$\,. The main black hole property changes are obtained as particle momenta. From the results, the black hole properties are related each others. Interestingly, the increase of the entropy and irreducible mass is dependent on $p^r$\, in the MP black holes. In the microscopic point of view, the particle radial momentum can increase the black hole energy degeneracy, the entropy, but the angular momenta cannot. The entropy changes can be classified into four cases according to the particle rotational directions and momenta at the horizon. If the particle rotates in the same direction as the black hole, the black hole mass and rotation energy, $M_{rot}$, are increased. For a particle with non-zero radial momentum, $p^r\,$, the entropy and irreducible mass are increased. It is interpreted as an irreversible process, since $\delta S_{BH}>0\,$. For a particle with zero radial momentum, the entropy and irreducible mass are not changed. In other words, $\delta S_{BH}\sim \delta M_{ir} =0\,$, which can be interpreted as a reversible process. If the particle rotates in the opposite direction of the black hole, the black hole rotational energy is decreased. For a particle with non-zero radial momentum, $p^r\,$, the entropy and irreducible mass still increase as in an irreversible process. For the specific case of a particle with zero radial momentum, the black hole mass is decreased. The entropy and irreducible mass are not changed as in a reversible process.

\section{Generalization to Multi Rotating Planes}
The effects of additional spin parameters can be described in $N$-dimensional MP black holes with arbitrary rotations. The metric is given in Boyer-Lindquist coordinates in $N$-dimensions\,\cite{MP,emp1,VPage1}\,,
\begin{eqnarray}
&&ds^2=-dt^2+\frac{U}{V-2M}+\frac{2M}{U}\left[dt-\sum_{i=1}^{n-\epsilon}a_i\mu^2_id\phi_i\right]^2+\sum_{i=1}^{n}(r^2+a_i^2)d\mu_i^2+\sum_{i=1}^{n-\epsilon}(r^2+a_i^2)\mu_i^2d\phi_i^2\,\\
&&U=r^{\epsilon}\sum_{i=1}^{n}\frac{\mu_i^2}{r^2+a_i^2}\prod_{j=1}^{n-\epsilon}(r^2+a_j^2)\,,\,\,F=r^2\sum_{i=1}^{n}\frac{\mu_i^2}{r^2+a_i^2}\,,\,\,V=r^{\epsilon-2}\prod_{i=1}^{n-\epsilon}(r^2+a_i^2)=\frac{U}{F}\,,\nonumber
\end{eqnarray}
where $n=[N/2]$. Coordinates $\mu_i$ are satisfied the constraint $\sum_{i=1}^{n}\mu_i^2=1\,$. The MP black holes are written differently in the cases of even and odd dimensions, and are distinguished by an $\epsilon$ value of 1 for even and 0 for odd cases. Notice that there is no rotation in a last direction of the even dimension cases.

We consider two spin values $a$ and $b$ such that the spin parameter, $a_i$, has a value of $a$ for $i=1\,,\,\,...\,,m\,,$ and $b$ for $m+j=m+1\,,\,\,...\,,m+p\,,$ where $m+p=n-\epsilon\,$. There are no restrictions for the odd dimension cases, but the value of $b$ is zero for even dimension cases. To get an explicit form of the inverse metric, the coordinates $\mu_i$ are also explicit notation that $\mu_i=\lambda_i\sin\theta$ for $i=1\,,\,\,...\,,m\,,$ and $\mu_{j+m}=\nu_j\cos\theta$ for $j=1\,,\,\,...\,,p\,$. In this case, the first-order derivative geodesics are obtained in \,\cite{VPage1} by the Hamilton-Jacobi method. The constraint is separated to,
\begin{eqnarray}
\sum_{i=1}^{m}=\lambda_i^2=1\,,\,\,\sum_{j=1}^{p}=\nu_j^2=1\,.
\label{eq:secdcons}
\end{eqnarray}
The coordinate system satisfying eq.~(\ref{eq:secdcons}) is
\begin{eqnarray}
\lambda_i=\left[\prod_{k=1}^{m-i}\sin\alpha_k\right]\cos\alpha_{m-i+1}\,,\,\,\nu_j=\left[\prod_{k=1}^{m-i}\sin\beta_k\right]\cos\beta_{p-j+1}\,.
\end{eqnarray}
In the coordinate system\,\cite{VPage1}, the inverse metric is given,
\begin{eqnarray}
g^{\mu\nu}\partial_{\mu}\partial_{\nu}&=&\left[-1-\frac{2MV}{U(V-2M)}\right]\partial_t\partial_t-\sum_{i=1}^n\frac{4MVa_i}{U(V-2M)(r^2+a_i^2)}\partial_t\partial_{\phi_i}+\frac{V-2M}{U}\partial_r\partial_r\\
&&+\sum_{i=1}^n\sum_{j=1}^n\left[\frac{1}{(r^2+a_i^2)\mu_i^2}\delta^{ij}-\frac{2MVa_i a_j}{U(V-2M)(r^2+a_i^2)(r^2+a_j^2)}\right]\partial_{\phi_i}\partial_{\phi_j}+\frac{1}{\rho^2}\partial_\theta\partial_\theta\nonumber\\
&&+\sum_{i=1}^{m-1}\frac{1}{(r^2+a_i^2)\sin^2\theta\left[\prod_{k=1}^{i-1}\sin^2\alpha_k\right]}\partial_{\alpha_i}\partial_{\alpha_i}+\sum_{i=1}^{p-1}\frac{1}{(r^2+a_i^2)\cos^2\theta\left[\prod_{k=1}^{i-1}\sin^2\beta_k\right]}\partial_{\beta_i}\partial_{\beta_i}\,,\nonumber
\end{eqnarray}
where $\rho^2$ is defined to $\rho^2= r^2+a^2\cos^2\theta+b^2\sin^2\theta\,$. For general spin parameters, the inverse metric has a additional term between spin parameters, which are not written generally for given dimensions and spins\,\cite{VPage1}\,, hence may not be integrable. The integrable geodesic equations\,\cite{VPage1} are obtained from the Hamilton-Jacobi action,
\begin{eqnarray}
S=\frac{1}{2}m^2\lambda-Et+\sum_{i=1}^{m}\Phi_i\phi_i+\sum_{j=1}^{p}\Psi_j\phi_{m+j}+S_r(r)+S_\theta(\theta)+\sum_{i=1}^{m-1}S_{\alpha_i}(\alpha_i)+\sum_{j=1}^{p-1}S_{\beta_j}(\beta_j)\,.
\end{eqnarray}
Hamilton-Jacobi equations are separated into four equations corresponding to $r\,,\,\,\theta\,,\,\,\alpha_i\,,\,\,\beta_j\,$,
\begin{eqnarray}
\label{eq:separate}
&&J_1^2=\sum_{i=1}^{m}\left[\frac{\Phi_i^2}{\lambda_i^2}+\frac{1}{\prod_{k=1}^{i-1}\sin^2\alpha_k}\left(\frac{dS_{\alpha_i}}{\alpha_i}\right)^2\right]\,,\,\,\,\,L_1^2=\sum_{i=1}^{p}\left[\frac{\Psi_i^2}{\nu_i^2}+\frac{1}{\prod_{k=1}^{i-1}\sin^2\beta_k}\left(\frac{dS_{\beta_i}}{\beta_i}\right)^2\right]\,,
\end{eqnarray}
\begin{eqnarray}
\mathcal{K}=&&-m^2r^2+E^2\left[r^2+\frac{2MZ}{r^2\Delta}\right]-\frac{\Delta Z}{r^\epsilon \Pi}\left(\frac{dS_r}{dr}\right)^2-\frac{4Ma(r^2+b^2)}{r^2\Delta}\sum_{i=1}^{m}E\Phi_i-\frac{4Mb(r^2+a^2)}{r^2\Delta}\sum_{i=1}^{p}E\Psi_i\nonumber\\
&&+\frac{2Ma^2(r^2+b^2)}{\Delta r^2 (r^2+a^2)}\sum_{i=1}^{m}\sum_{j=1}^{m}\Phi_i\Phi_j+\frac{2Mb^2(r^2+a^2)}{\Delta r^2 (r^2+b^2)}\sum_{i=1}^{p}\sum_{j=1}^{p}\Psi_i\Psi_j+\frac{4Mab}{\Delta r^2}\sum_{i=1}^{m}\sum_{j=1}^{p}\Phi_i\Psi_j\nonumber\\
&&-\frac{r^2+b^2}{r^2+a^2}J_1^2-\frac{r^2+a^2}{r^2+b^2}L_1^2\,,\nonumber\\
\mathcal{K}=&&(m^2-E^2)(a^2\cos^2\theta+b^2\sin^2\theta)+\left(\frac{dS_\theta}{d\theta}\right)^2+\cot^2\theta J_1^2+\tan^2\theta L_1^2\,,\nonumber
\end{eqnarray}
in which $J_1$, $L_1$, and $\mathcal{K}$ are separate constants. Note that the additional definitions are,
\begin{eqnarray}
\label{eq:NArbtconst}
\Delta=V-2M\,,\,\,\Pi=(r^2+a^2)^m(r^2+b^2)^{p-\epsilon}\,,\,\,Z=(r^2+a^2)(r^2+b^2)\,.
\end{eqnarray}
All of geodesic equations are obtained from eq.~(\ref{eq:separate}). There are geodesics for $\rho$ and $\theta$ coordinates to determine the energy of the particle at the specific momenta and locations. The geodesic equations are given,
\begin{eqnarray}
\label{eq:geo}
&&\rho^2\frac{dr}{d\lambda}=\frac{\Delta Z}{\Pi r^\epsilon}\sqrt{R}\,,\,\,\,\,\rho^2\frac{d\theta}{d\lambda}=\sqrt{\Theta}\,,
\end{eqnarray}
where the functions $R$ and $\theta$ are obtained,
\begin{eqnarray}
\frac{\Delta Z}{\Pi r^\epsilon}R=&&-\mathcal{K}+(E^2-m^2)r^2+\frac{2MZ}{r^2\Delta}E^2-\frac{4Ma(r^2+b^2)}{r^2\Delta}\sum_{i=1}^{m}E\Phi_i-\frac{4Mb(r^2+a^2)}{r^2\Delta}\sum_{i=1}^{p}E\Psi_i\\
&&+\frac{2Ma^2(r^2+b^2)}{\Delta r^2 (r^2+a^2)}\sum_{i=1}^{m}\sum_{j=1}^{m}\Phi_i\Phi_j+\frac{2Mb^2(r^2+a^2)}{\Delta r^2 (r^2+b^2)}\sum_{i=1}^{p}\sum_{j=1}^{p}\Psi_i\Psi_j+\frac{4Mab}{\Delta r^2}\sum_{i=1}^{m}\sum_{j=1}^{p}\Phi_i\Psi_j\,,\nonumber\\
&&-\frac{r^2+b^2}{r^2+a^2}J_1^2-\frac{r^2+a^2}{r^2+b^2}L_1^2\,,\nonumber\\
\Theta=&&\mathcal{K}-(m^2-E^2)(a^2\cos^2\theta+b^2\sin^2\theta)-\cot^2\theta J_1^2-\tan^2\theta L_1^2\,.\nonumber
\end{eqnarray}
The particle energy obtained from eq. (\ref{eq:geo}) is given,
\begin{eqnarray}
\alpha E^2 + \beta E + \gamma = 0\,,
\end{eqnarray}
\begin{eqnarray}
&\alpha=&\rho^2+\frac{2MZ}{r^2\Delta}\,\,\nonumber\\
&\beta=&-\frac{4Ma(r^2+b^2)}{r^2\Delta}\sum_{i=1}^{m}\Phi_i-\frac{4Mb(r^2+a^2)}{r^2\Delta}\sum_{i=1}^{p}\Psi_i\,,\nonumber\\
&\gamma=&-\rho^4\left({p^\theta}^2+{p^r}^2\frac{\Pi r^\epsilon}{\Delta Z}\right)-\rho^2 m^2-\cot^2\theta J_1^2-\tan^2\theta L_1^2+\frac{4Mab}{\Delta r^2}\sum_{i=1}^{m}\sum_{j=1}^{p}\Phi_i\Psi_j\nonumber\\
&&+\frac{2Ma^2(r^2+b^2)}{\Delta r^2 (r^2+a^2)}\sum_{i=1}^{m}\sum_{j=1}^{m}\Phi_i\Phi_j+\frac{2Mb^2(r^2+a^2)}{\Delta r^2 (r^2+b^2)}\sum_{i=1}^{p}\sum_{j=1}^{p}\Psi_i\Psi_j-\frac{r^2+b^2}{r^2+a^2}J_1^2-\frac{r^2+a^2}{r^2+b^2}L_1^2\,,\nonumber
\end{eqnarray}
There is a minimum energy, $E_{min}$, since $\alpha$ is positive outside the horizon. The momenta and mass of the black hole are increased as much as that of the absorbed particle. The particle energy defined by its momenta at the horizon is obtained,
\begin{eqnarray}
\label{NArbtE1}
E=\frac{a}{r_h^2+a^2}\sum_{i=1}^{m}\Phi_i+\frac{b}{r_h^2+b^2}\sum_{j=1}^{p-\epsilon}\Psi_j+\frac{|\rho^2 p^r r_h^2|}{(r_h^2+a^2)(r_h^2+b^2)}\,,
\end{eqnarray}
where its minimum energy is the summation of $E_{min}$ in eq.~(\ref{eq:NminE}) for given rotating planes. The minimum energy is also achieved when $p^r=0$\,. The black hole rotates in all planes, so all of the particle angular momenta can increase the black hole angular momenta. The effects of particle and black hole angular momenta are separated from each others. The information about dimensionality and rotations is only included in $r_h$. The particle energy, $E$, and angular momenta, $\Phi_i$s and $\Psi_j$s match the changes of the black hole mass $M_B$ and angular momenta $J_{a_i}$ and $J_{b_j}$\,. The mass and momenta of a MP black hole absorbing the particle are changed to,
\begin{eqnarray}
\label{NArbtE2}
\delta M_B= \frac{a}{r_h^2+a^2}\sum_{i=1}^{m}\delta J_{a_i}+\frac{b}{r_h^2+b^2}\sum_{j=1}^{p-\epsilon}\delta J_{b_j}+\frac{|\rho^2 p^r r_h^2|}{(r_h^2+a^2)(r_h^2+b^2)}\,,
\end{eqnarray}
where $J_{a_i}$ and $J_{b_j}$ are angular momenta with respect to spin parameters $a$ and $b$. The black hole mass is always increased since the changes are positive for every bit of absorbed matter. The irreducible mass, $M_{ir}$, and its change, $\delta M_{ir}$, are obtained explicitly,
\begin{eqnarray}
\label{NArbitMir1}
&&M_{ir}=\sqrt{\frac{(r_h^2+a^2)^m(r_h^2+b^2)^{n-m-\epsilon}}{r_h^{1-\epsilon}}}=\sqrt{2Mr_h}\,,\\
&&\delta M_{ir}=\frac{8\pi r_h^3|p^r|\rho^2(r_h^{\epsilon-1}(r_h^2+a^2)^m(r_h^2+b^2)^{n-m-\epsilon})^{-\frac{1}{2}}}{\Omega_{N-2}(r_h^2(2b^2(-1+m)+r_h^2(-2+2n-\epsilon))-a^2(2b^2+r_h^2(2+2m-2n+\epsilon)))}\geq 0\,,\nonumber
\end{eqnarray}
where $\epsilon=1$ and $b=0$ gives even cases. The extended irreducible mass is also written as the mass parameter and horizon. It includes the Kerr black hole, 5-dimensional, and a single rotating MP black hole case. The rotation plane added does not depend on irreducible mass. Using eq.~(\ref{NArbtE2}) and (\ref{NArbitMir1}), the horizon and spin parameter changes are obtained as particle momenta,
\begin{eqnarray}
&&\delta a=\frac{1}{M_B}\left[\frac{N-2}{2}\Phi_i-\frac{a^2}{r_h^2+a^2}\sum_{i=1}^{m}\Phi_i+\frac{ab}{r_h^2+b^2}\sum_{j=1}^{p-\epsilon}\Psi_j+\frac{|\rho^2 p^r r_h^2|}{(r_h^2+a^2)(r_h^2+b^2)}\right]\,,\\
&&\delta b=\frac{1}{M_B}\left[\frac{N-2}{2}\Psi_i-\frac{ab}{r_h^2+a^2}\sum_{i=1}^{m}\Phi_i+\frac{b^2}{r_h^2+b^2}\sum_{j=1}^{p-\epsilon}\Psi_j+\frac{|\rho^2 p^r r_h^2|}{(r_h^2+a^2)(r_h^2+b^2)}\right]\,,\nonumber\\ 
&&\delta r_h = -\frac{a}{r_h^2+a^2}\sum_{i=1}^{m}\delta \Phi_i-\frac{b}{r_h^2+b^2}\sum_{j=1}^{p-\epsilon}\Psi_j\nonumber\\
&&+\frac{1}{M_B}\left(\frac{(N-2)r_h^3\rho^2)^{-\frac{1}{2}}}{r_h^2(2b^2(-1+m)+r_h^2(-2+2n-\epsilon))-a^2(2b^2+r_h^2(2+2m-2n+\epsilon))}-\frac{r_h^3\rho^2}{(r_h^2+a^2)(r_h^2+b^2)}\right)|p^r|\,.\nonumber
\end{eqnarray}
The surface area of the MP black hole is given,
\begin{eqnarray}
&&\mathcal{A}=\frac{\Omega_{N-2}}{\kappa}M\left(N-3-\sum_{i=1}^{n-\epsilon}\frac{2a_i^2}{r_h^2+a_i^2}\right)\,,\\&&\kappa=\lim_{r\rightarrow r_h}\frac{\partial_r(r^{2}V)-4Mr}{4Mr^2}\,\,(odd\,\,N)\,,\,\,\,\,\kappa=\lim_{r\rightarrow r_h}\frac{\partial_r(rV)-2Mr}{4Mr}\,\,(even\,\,N)\nonumber\,,
\end{eqnarray}
in which $\kappa$ is the surface gravity, and $\kappa=0$ is a extremal case. Interestingly, the area can be written simply to one expression without dimensionality dependency and related to irreducible mass,
\begin{eqnarray}
\mathcal{A}=2\Omega_{N-2}Mr_h=\Omega_{N-2}M_{ir}^2\,.
\end{eqnarray}
The Bekenstein-Hawking entropy and its change is obtained,
\begin{eqnarray}
\label{NarbitS}
S_{BH}=\frac{1}{4}\Omega_{N-2}M_{ir}^2\,,\,\,\,\,\delta S_{BH}=\frac{4\pi r_h^3|p^r|\rho^2}{r_h^2(2b^2(-1+m)+r_h^2(-2+2n-\epsilon))-a^2(2b^2+r_h^2(2+2m-2n+\epsilon))}\geq 0\,.
\end{eqnarray}
Note that the entropy and irreducible mass have a one-to-one correspondence. The particle absorption can change the entropy and irreducible mass, since it affects the mass of the black hole, $M_B\,$. The changes in entropy and irreducible mass are only proportional to the particle radial momentum $p^r$ at the horizon from eq.~(\ref{NArbitMir1}) and (\ref{NarbitS}). It can be interpreted that the particle radial momentum increases the black hole energy degeneracy, the entropy. The particle angular momenta do not change the energy degeneracy, though it can change the black hole rotation energy.

In the analysis, we understand that particle absorption induces thermodynamic transformations of the black hole. The process can be systematically described for all dimensions and spins. The changes of the properties of the MP black holes, including that of Kerr black holes, can be classified into four cases according to the particle rotational directions and momenta at the horizon. If the particle rotates in the same direction as the black hole, the black hole mass is increased. For a particle with a non-zero radial momentum, the entropy and irreducible mass are increased. It is interpreted as an irreversible process, since $\delta S_{BH}>0\,$. From the increasing $M_{ir}\,$ in eq.~(\ref{eq:delta_Mir}) and (\ref{NArbitMir1})\,, the change of the black hole mass can be shown to be inversely proportional to that of the horizon, $\frac{dr_h}{dM_B}>-\frac{r_h}{M_B}$. For a single case, the horizon changes are,
\begin{eqnarray}
|p^r|\geq\frac{((N-3)r_h^2+(N-5)a^2)}{r_h^2+3a^2}\frac{aL}{\Sigma}\,\rightarrow\,\delta r_h\geq 0\,,\,\,|p^r|\leq\frac{((N-3)r_h^2+(N-5)a^2)}{r_h^2+3a^2}\frac{aL}{\Sigma}\,\rightarrow\,\delta r_h\leq 0\,.
\end{eqnarray}
The changes of the black hole spin parameters, $\delta a$s and $\delta b$s are also dependent on the values of the particle momenta. Explicit expressions for a single rotating case are,
\begin{eqnarray}
|p^r|\leq\frac{((N-2)r_h^2+(N-4)a^2)L}{a\Sigma}\,\rightarrow\,\delta a\geq 0\,,\,\,|p^r|\geq\frac{((N-2)r_h^2+(N-4)a^2)L}{a\Sigma}\,\rightarrow\,\delta a\leq 0\,.
\end{eqnarray}
The black hole rotation energy increases because $\delta M_B-\delta M_{ir} >0\,$, which comes from the particle angular momenta. For the specific case of a particle with zero radial momentum, the entropy and irreducible mass are unchanged. In the other words, $\delta S_{BH}\sim \delta M_{ir} =0\,$. It can be interpreted as a reversible process. Also, black hole property changes become simple. The change of the black hole mass is inversely proportional to that of the horizon, $\frac{dr_h}{dM_B}=-\frac{r_h}{M_B}$, so the black hole horizon becomes smaller. The black hole rotational energy increases due to the bigger $M_B$ and spin parameters. In addition, the MP black holes with a single rotation always have a bigger spin parameter and smaller horizon after absorption. 

If the particle rotates in the opposite direction to the black hole, the black hole mass can decrease. The black hole mass increases when the radial momentum is bigger than $\frac{a|L|}{\Sigma}$, and decreases when smaller than $\frac{a|L|}{\Sigma}\,$. For a particle with a non-zero radial momentum, $p^r\,$, the entropy and irreducible mass still increase as an irreversible process. From the increasing $M_{ir}\,$, the change of the black hole mass is inversely proportional to that of the horizon, $\frac{dr_h}{dM_B}>-\frac{r_h}{M_B}$, and the changes of the black hole spin parameters, $\delta a$s and $\delta b$s are dependent on the values of the particle momenta. For a single rotating case, a spin parameter always decreases from a loss of rotation energy. The horizon size changes are dependent on spacetime dimensions. For $N$=4 Kerr case, the horizon becomes bigger than before when $|p^r|\geq\frac{(r_h^2-a^2)aL}{(r_h^2+3a^2)\Sigma}\,$ and smaller when $|p^r|<\frac{(r_h^2-a^2)aL}{(r_h^2+3a^2)\Sigma}\,$. For $N\geq5$ black holes, the horizon becomes bigger for a non-zero radial momentum. The decreasing black hole mass only comes from the decreasing black hole rotational energy, because $\delta M_B-\delta M_{ir} <0\,$, and $\delta M_{ir} >0\,$. For the specific case of a particle with zero radial momentum, $\delta S_{BH}\sim \delta M_{ir} =0\,$ as a reversible process. The change of the black hole mass is inversely proportional to that of the horizon, $\frac{dr_h}{dM_B}=-\frac{r_h}{M_B}$, so the black hole horizon becomes larger. The black hole rotational energy is decreased by the smaller $M_B$ and spin parameters\, for arbitrary dimensions and rotating planes. For the case of $N$=4, it is related to the Penrose process\,\cite{pen} in the Kerr black hole.

\section{Myers-Perry Black Hole Phases}
There are special cases in MP black holes. To show what happens in the black hole in our universe or higher dimensional spacetimes, we consider restricted examples and change their properties by throwing a particle. We consider the MP black hole with a single rotation. Since a black hole absorbs a particle, then it changes its mass and angular momenta, the black hole mass and spin parameter are different to each stages, but these are evolved from one black hole. Mainly, the horizon sizes are discussed to show the topological phase changes. A $N$=4 MP black hole with a single rotation describes a Kerr black hole. When a spin parameter is zero, it goes to a Schwarzschild black hole and has a horizon, $r_h=2M$. If the spin is fixed at $a=M$\,, it is called the extremal Kerr black hole. The extremal Kerr black hole still maintains its properties by $\delta M_B \geq \delta a$\,\cite{Bardeen}\,. If the black hole starts as $a_1\ll M_1$ and absorbs a particle of same rotating to the black hole,
\begin{eqnarray}
a_1\ll1\,,\,\, \delta r_h \leq 0\,\,\rightarrow\,\,|p^r|\leq \frac{a_1}{\Sigma} L\,,\,\,\delta r_h > 0\,\,\rightarrow\,\,|p^r|>\frac{a_1}{\Sigma} L\,,
\end{eqnarray}
where $a_1$ is very small values, so the horizon size almost becomes bigger than before absorption. After absorbing particles, the black hole mass increases and a spin parameter becomes $a_2 < M_2$. The horizon change is written,
\begin{eqnarray}
a_2 < M_2\,,\,\, \delta r_h \leq 0\,\,\rightarrow\,\,|p^r|\leq \frac{a_2(r_h^2-a_2^2)}{\Sigma(r_h^2+3a_2^2)} L\,,\,\,\delta r_h > 0\,\,\rightarrow\,\,|p^r|> \frac{a_2(r_h^2-a_2^2)}{\Sigma(r_h^2+3a_2^2)} L\,,
\end{eqnarray}
in which $r_h^2-a_2^2<0$ in $a_2 < M_2$, so the horizon size increases for any particles having a non-zero radial momentum. However, if the black hole catches zero-radial momentum particle, ideally, the horizon can be made smaller. It finally achieves $a_3=M_3$. Note that the black hole rotation energy increase causes a bigger mass of that. In the extremal Kerr black hole,
\begin{eqnarray}
a_3 = M_3\,,\,\, \delta r_h \leq 0\,\,\rightarrow\,\,|p^r|\leq0\,,\,\,\delta r_h > 0\,\,\rightarrow\,\,|p^r|>0\,,
\end{eqnarray}
where $|p^r|<0$ cannot be achieved in our real world. Thus, the non-zero radial momentum particle always makes the horizon size bigger, and breaks extremality. In addition, the black hole catches zero-radial momentum particle, ideally, the horizon increases as much as mass increase, $\delta r_h=\delta M_3 >0$. It conserves the extremality. From this Kerr black hole evolution, making a smaller horizon is very difficult task in nature. The extremal case is too hard to maintain. It is broken just one non-zero radial particle.

The extremality problems can be extended to the MP black holes in the upper bound cases. The spin upper bound with a single rotation is quite different. For the $N=$5 case, the horizon is written as $\sqrt{2M-a^2}$ and disappears at $a=\sqrt{2M}$. It is a naked ring singularity. Analyzing eq.~(\ref{eq:delta_Mir})\,, the thermodynamic properties are changed and classified into $0\leq a \leq \sqrt{M}$ and $\sqrt{M}\leq a \leq \sqrt{2M}$\,. Now, the black hole starts with $a_1\ll 1$ and mass $M_1$\,. In the case of absorbing a same rotating particle, the horizon changes,
\begin{eqnarray}
a_1\ll1\,,\,\, \delta r_h \leq 0\,\,\rightarrow\,\,|p^r|\leq \frac{2a_1}{\Sigma} L\,,\,\,\delta r_h > 0\,\,\rightarrow\,\,|p^r|> \frac{2a_1}{\Sigma} L\,,
\end{eqnarray}
where $a_1$ is small variable and $r_h\sim \sqrt{2M}\,$. The horizon is bigger than before by any non-zero radial momentum particle. Absorbing particles, the black hole has bigger mass $M_2$, a spin is in a range of $0<a_2<\sqrt{M}$. The horizon now can be smaller by,
\begin{eqnarray}
0<a_2<\sqrt{M}\,,\,\, \delta r_h \leq 0\,\,\rightarrow\,\,|p^r|\leq \frac{a_2(2r_h^2)}{\Sigma(r_h^2+3a_2^2)} L\,,\,\,\delta r_h > 0\,\,\rightarrow\,\,|p^r|> \frac{a_2(2r_h^2)}{\Sigma(r_h^2+3a_2^2)} L\,.
\end{eqnarray}
In consideration about $M_2>M_1$ and $a_2>a_1$, the radial momentum which can make the horizon smaller is also broader than before. This range is the largest when $a=\sqrt{M_3}\,$ after absorption. In this case, the horizon behavior is obtained,
\begin{eqnarray}
a_3=\sqrt{M_3}\,,\,\, \delta r_h \leq 0\,\,\rightarrow\,\,|p^r|\leq \frac{M_3}{2\Sigma} L\,,\,\,\delta r_h > 0\,\,\rightarrow\,\,|p^r|> \frac{M_3}{2\Sigma} L\,,
\end{eqnarray}
where $r_h=\sqrt{M_3}\,$. A spin parameter increases bigger than this value. The decreasing horizon size is more difficult than before, but it can be possible. After absorbing more particles, the black hole reaches to $M_4>M_3$ and $\sqrt{M_4}<a_4<\sqrt{2M_4}\,$. The horizon behaves as,
\begin{eqnarray}
\sqrt{M_4}<a_4<\sqrt{2M_4}\,,\,\, \delta r_h \leq 0\,\,\rightarrow\,\,|p^r|\leq \frac{a_4(2r_h^2)}{\Sigma(r_h^2+3a_4^2)} L\,,\,\,\delta r_h > 0\,\,\rightarrow\,\,|p^r|> \frac{a_4(2r_h^2)}{\Sigma(r_h^2+3a_4^2)} L\,.
\end{eqnarray}
With taking small $p^r$ particles, the black hole of mass $M_5$ slowly converges to the point of $a_5\sim\sqrt{2M_5}$. However, the making smaller horizon is almost impossible in $a_5\sim\sqrt{2M_5}\,$. It can be shown as,
\begin{eqnarray}
a_5\sim \sqrt{2M_5}\,,\,\, \delta r_h \leq 0\,\,\rightarrow\,\,|p^r|\leq 0\,,\,\,\delta r_h > 0\,\,\rightarrow\,\,|p^r|> 0\,.
\end{eqnarray}
Note that $r_h\sim 0$. Therefore, the horizon size increase for any particle absorption, and watching a naked singularity is avoidable even though 5-dimensional black holes. In fixed a black hole mass, changing a black hole spin causes decreasing irreducible mass, that the entropy is decreased. A observer sees the black hole mass as a sum of $M_{ir}+M_{rot}\,$. So, it is not a physical evolution of a black hole, However, in our analysis, the black hole evolves by $\delta M_B$ and $\delta J$ carried from a particle. This process satisfies thermodynamic 2nd law as $\delta S_{BH}>0$. The black hole mass increases also in the case of bigger rotation energy, even if irreducible mass is not increased. It can be more physical evolution of the black hole in the universe. From the reason that the black hole increases its mass as much as the particle energy, the smaller horizon is allowed in very small ranges. The small range is disappeared in a small horizon. Therefore, the disappearing horizon in 5-dimensional MP black hole is not allowed in the real universe. Note that, for the opposite rotating cases, the horizon size becomes larger for $p^r<\frac{2r_h^2 aL}{\Sigma(a^2-r_h^2)}$ in $0<a<\sqrt{M}$ and any values of momenta in $\sqrt{M}<a<\sqrt{2M}$ from eq.~(\ref{eq:delta_Mir})\,, so it also can avoid from the naked singularity.

For $N\geq 6\,$, there exists no upper bound. Since these have a horizon for all values of spin, so there is no restriction on the spin value. If a large spin parameter is achieves as $a\gg r_h$\,, it is called ultra-spinning black hole. Since its topology is vast thin disk, it is also called black membrane. To make a ultra-spinning case in $N\geq6$, the momenta of a falling particle are in the range which can achieve $\delta r_h <0$ and $\delta a> 0$ at the same time. The satisfying particle radial momentum is given,
\begin{eqnarray}
\delta a \geq 0 \,\,\rightarrow\,\,|p^r|\leq\frac{(N-2)r_h^2+(N-4)a^2}{a^2}\left(\frac{aL}{\Sigma}\right)\,,\,\,\delta r_h \leq 0 \,\,\rightarrow\,\,|p^r|\leq\frac{(N-3)r_h^2+(N-5)a^2}{r_h^2+3a^2}\left(\frac{aL}{\Sigma}\right)\,,
\end{eqnarray}
The necessary condition is for $\delta r_h\,$, so smaller horizon also means bigger a spin value. Starting with a $N\geq6$ MP black hole of $a\sim 0$ and mass $M_1$, the property changes is given,
\begin{eqnarray}
a_1\ll1\,,\,\,\delta a \geq 0 \,\,\rightarrow\,\,|p^r|\leq(N-2)r_h^2\left(\frac{L}{a_1\Sigma}\right)\,,\,\,\delta r_h \leq 0\,\,\rightarrow\,\,|p^r|\leq (N-3) \left(\frac{a_1 L}{\Sigma}\right)\,.
\end{eqnarray}
Since $a\ll 1$, $\delta a > 0$ is possible for any particle momenta and $\delta r_h <0$ is achieved very small radial momenta. The black hole increases its mass to $M_2\,$. The changes are,
\begin{eqnarray}
\delta a_2 \geq 0 \,\,\rightarrow\,\,|p^r|\leq\frac{(N-2)r_h^2+(N-4)a_2^2}{a_2^2}\left(\frac{a_2 L}{\Sigma}\right)\,,\,\,\delta r_h \leq 0 \,\,\rightarrow\,\,|p^r|\leq\frac{(N-3)r_h^2+(N-5)a_2^2}{r_h^2+3a_2^2}\left(\frac{a_2L}{\Sigma}\right)\,,
\end{eqnarray}
in which the ranges of radial momentum are broader than $a_1\ll 1$ case. The particle in these ranges makes the black hole thin and vast. However, the range becomes very limited, again. Absorbing this limited momenta particles, the black hole of the mass $M_3$ is in the stage at $a\gg r_h$\,. The changes are described as
\begin{eqnarray}
\delta a_3 \geq 0 \,\,\rightarrow\,\,|p^r|\leq(N-4)\left(\frac{a_3 L}{\Sigma}\right)\,,\,\,\delta r_h \leq 0 \,\,\rightarrow\,\,|p^r|\leq\frac{(N-5)}{3}\left(\frac{a_3L}{\Sigma}\right)\,,
\end{eqnarray}
where $a\gg r_h$, so increasing a spin is impossible any more after reaching to the black membrane. However, it is possible, ideally. Shortly, For the opposite rotating cases, the horizon size becomes smaller when $|p^r|>\frac{(N-3)r_h^2+(N-5)a^2}{(4-N)r_h^2+(6-N)a^2}aL$ from eq.~(\ref{eq:delta_Mir})\,. The $N\geq6$ MP black holes which absorb many particles become $a\gg r_h$\,; the ultra spinning case.

Next, the MP black holes with arbitrary rotations are considered, briefly. For $N=$5 with arbitrary rotations, the upper bounds of the spins are given as $a+b=\sqrt{2M}$ for $a,b>0$. After absorption of the particle, the changes are rewritten as $\delta \sqrt{2M}\geq\delta a+\delta b$ from eq.~(\ref{NArbtE2}) in which the equality is satisfied at zero radial momentum. Thus, the black hole is still in the spin-bound case by the zero radial momentum particle absorption, but the non-zero radial momentum particle absorption breaks it into two horizons and increases $M_{ir}$\,. For $N\geq 6$ cases, the existence of the spin upper bound cases is dependent on the number of spins and their values\,\cite{emp1}. The spin upper bounds are determined by algebraically solving an equation of degree 2($N$-3). It is generally difficult to do, however we expect that the changes of the MP black hole properties will follow our results explained above.

\section{Interpretation of Particle Absorption in a Dual CFT}
Since a particle carries momenta, it changes the black hole energy in microscopically. Using a progress of a dual CFT description corresponding to extremal Kerr black hole\,\cite{kerrCFT}, microscopic meanings of the particle absorption are shown as a dual CFT point of view. A single rotating MP black hole at $N$=4 becomes Kerr black hole. To apply to quantum theory, the angular momentum $J$ is quantized as $J=\hbar j$\,. The Hawking temperature of a Kerr black hole is,
\begin{eqnarray}
T_H=\frac{\hbar (r_h-M)}{4\pi M r_h}\,.
\end{eqnarray}
Note that $\hbar$ set to 1 for simplicity. The Kerr/CFT correspondence is satisfied at extremal Kerr black hole. In MP black hole coordinate system, the extremal condition gives $a=r_h=M\,$, and the entropy and relation of dual CFT are written,
\begin{eqnarray}
\label{expJ}
S_{BH}=2\pi J=\frac{\pi^2}{3}c_L T_L=S_{micro}\,.
\end{eqnarray}
Rewriting these relations as a irreducible mass gives,
\begin{eqnarray}
\label{expMir}
S_{BH}=\pi M_{ir}^2\,,\,\,J=\frac{M_{ir}^2}{2}\,,
\end{eqnarray}
which is same to eq.~(\ref{singleS1}). The descriptions of eq.~(\ref{expJ}) and (\ref{expMir}) have important differences. If the zero radial momentum particle falls to the black hole with opposite rotating direction, the angular momentum can decrease. For eq.~(\ref{expJ}) case, $\delta J<0$ cause to $\delta S_{BH}<0$, and for eq.~(\ref{expMir}), $\delta S_{BH}=0$ from eq.~(\ref{eq:delta_Mir}). The description of eq.~(\ref{expJ}) is satisfied each $a=r_h=M$ Kerr black hole, so it goes to empty spacetime at $J=0$, and the irreducible mass is implicitly changed when we control $J\,$. However, the description of eq.~(\ref{expMir}) has the concept of irreducible mass. We can change $J$ by a particle without irreducible mass change. When $J=0$, it goes to Schwarzschild black hole. To consider the particle absorption which maintain extremal Kerr black hole after the absorption, the reversible process is given to the extremal Kerr black hole. From eq.~(\ref{deltaarh1}), the zero radial momentum particle changes the black hole mass, spin parameter, and horizon as,
\begin{eqnarray}
\delta M= \delta a = \delta r_h=\frac{1}{2M}\,.
\end{eqnarray}
However, the irreducible mass and entropy does not change from eq.~(\ref{eq:delta_Mir}) and (\ref{singleS1}). The central charge $c_L$ is also written,
\begin{eqnarray}
c_L=12J=6M_{ir}^2\,,
\end{eqnarray}
Since irreducible mass is not changed in the case, the central charge is also invariant from this process. In a dual CFT point of view, the boundary conditions of the CFT requiring is not affected by the absorption, so the central charge is not changed, too. To comment the change of left temperature, there are need to define the quantum vacuum for extreme Kerr. It is also obtained as the Frolov-Thorne vacuum in \cite{kerrCFT}. The left and right temperatures with matching momentum eigenvalues are,
\begin{eqnarray}
T_L=\frac{r_h-M}{2\pi(r_h-a)}\,,\,\,T_R=\frac{r_h-M}{2\pi(\lambda r_h)}\,,\,\,n_L=m\,,\,\,n_R=\frac{1}{\lambda}(2M\omega-m)\,,
\end{eqnarray}
where $m$ and $\omega$ are quantum number and asymptotic energy of a field live outside of outer Kerr horizon. After the absorption, their changes are written,
\begin{eqnarray}
&&\delta T_L=\frac{\delta r_h-\delta M}{2\pi(r_h-a)}-\frac{(r_h-M)(\delta r_h-\delta a)}{2\pi(r_h-a)^2}\,,\,\,\delta T_R=\frac{\delta r_h-\delta M}{2\pi(\lambda r_h)}-\frac{(r_h-M)\delta r_h}{2\pi(\lambda r_h^2)}\,,\\
&&\delta n_L=0\,,\,\,\delta n_R=\frac{2\delta M\omega}{\lambda}\,,
\end{eqnarray}
where $m$ and $\omega$ are quantities of a field, so not changed. For extremal condition, $a=r_h=M\,$, these are,
\begin{eqnarray}
T_L=\frac{1}{2\pi}\,,\,\,T_R=0\,,\,\,n_L=m\,,\,\,n_R=\frac{1}{\lambda}(2M\omega-m)+\frac{\omega}{\lambda M}\,,
\end{eqnarray}
where the right momentum number is only changed. Therefore, the particle momenta affect to right momentum number. In this example, a particle with a zero radial momentum changes a extremal Kerr black hole to maintain extremality. This process increases $n_R$ for same rotating cases and decreases for opposite rotating cases. The temperature is sill $T_L$, so the microscopic entropy is also not changed. The Kerr black hole satisfies Kerr/CFT correspondence.

\section{Summary and Discussions}
In this paper, we investigate the black hole entropy using test particle absorption. We formulate the entropy change in terms of the particle momenta, and generalize the particle absorption to higher-dimensional MP black holes using the integrable structure. Using these results, we understand how one black hole changes its properties in a particle absorption. Also, this process gives relevant way for controlling black hole variables. If one changes a black hole mass or spin on the metric, it can be cause to decrease the entropy. That is not happened in our universe. However, if the mass or spin changes along to a particle momenta, the controlling variables can be done in satisfaction of the thermodynamic laws. In addition, it is what the astrophysical black hole does in the real world. By more realistic this method, we can trace MP black hole evolutions due to the particle absorption. Microscopical interpretation in limited cases is attempted using extremal Kerr/CFT correspondence. The restriction comes from maintaining extremality, so we discuss about the microscopic changes in reversible process. Since the entropy does not change, it cannot affect to physical vacuum, but it changes momentum number and temperature which are not physically favorable direction. First, we consider the particle absorption in N-dimensional MP black holes with a single rotation. The geodesics are obtained to find the particle energy constructed as specific locations and momenta. The black hole mass is increased as much as the particle energy at the horizon, depending on the particle angular and radial momenta. Using this, we construct the explicit forms of irreducible mass and its change. The irreducible mass is a quantity which increases due to the absorption. The square of the irreducible mass has a one-to-one correspondence to the Bekenstein-Hawking entropy. The change of the irreducible mass is described in terms of the radial momentum of the falling particle. Using one-to-one correspondence to the entropy, the change of the entropy is dependent on the radial momentum of the particle. In this case, the information about the dimensionality is in the horizon. The particle momenta in the additional spatial dimensions do not affect to the black hole charges. Secondly, we generalize the MP black holes with arbitrary rotations. In this work, we restrict the spin parameters to two values: $a$ of $m$ and $b$ of $p$\,, and the total number of spins are $m+p=n-\epsilon\,$, where $n=[\frac{N}{2}]$\,. For even $N$ cases, $\epsilon=1$ and $b=0$\,. For odd $N$ cases, $\epsilon=0$ and there is no restriction on the values of $a$ and $b$\,. In the analysis, the particle energy is written as the particle angular momenta in the black hole rotating planes and radial momentum. In the absorption, the changes of the entropy and irreducible mass are still dependent on the particle radial momentum. The entropy has a one-to-one correspondence to the squared irreducible mass. The information about the number of the black hole spins is only in the horizon.

In the analysis, the changes of the MP black hole properties can be systematically described for all dimensions and spins. The black hole mass is changed by the particle energy. The particle energy is determined by the particle radial momentum at the outer horizon and angular momenta in the black hole rotating planes. Since the black hole mass is divided into the irreducible mass and rotation energy, the change of irreducible mass is obtained by removing the contribution of the particle rotation energy. The square of the irreducible mass is proportional to the entropy. In the absorption, the black hole properties are changed by charges carried from a particle in the ranges of conserving thermodynamic 2nd law. The changes are exactly obtained as a particle momenta. Each property changes is differently dependent on particle radial and angular momenta. Thus, the physical interpretation should be different along the origin of changes. For example, the same black hole mass increase can be from irreducible mass or rotation energy increase, but the changes caused from this case affect to other changes and are shown to be different. Especially, the change of irreducible mass is only proportional to the particle radial momentum, so therefore the change of the entropy is only dependent on the particle radial momentum. It can be interpreted in the microscopic point of view that the particle radial momentum can increase the black hole energy degeneracy, entropy, but the angular momenta cannot. We can explain the thermodynamic processes induced by the particle absorption in the MP black holes including the Kerr black hole. The irreversible process occurs in the case of non-zero particle radial momentum. Here the entropy and irreducible mass are increased. The change of the black hole mass is inversely proportional to the change of the horizon. It is written as an inequality by $\frac{dr_h}{dM_B}>-\frac{r_h}{M_B}$\,. If the rotational direction of the particle and black hole are the same, the black hole mass and rotational energy are increased. If the particle and black holes rotate in opposite directions, then the black hole rotational energy is decreased. The black hole mass is however dependent on the values of the irreducible mass and rotational energy. The reversible process occurs in the case of zero particle radial momentum, and the entropy and irreducible mass are not changed. The change of the black hole mass is still inversely proportional to the change of the horizon. In this case, it is written as an equality by $\frac{dr_h}{dM_B}=-\frac{r_h}{M_B}$\,. If the directions of rotation of the particle and black hole are the same, the black hole mass, rotation energy and spin parameters are increased, and the horizon size is decreased. If the direction of rotation of the particle and black hole are opposite, the black hole mass, rotation energy, and spin parameters decrease, but the horizon size is increased.

We explain the phase changes of the MP black holes properties at the spin upper bounds for one-spin cases of all dimensions and 5-dimensional two-spins. First, the one-spin black hole property changes are dependent on the dimensionality and classified with three cases, $N=$4, 5, and over 6 dimensions\,. We trace the these black hole evolutions by a particle absorption. For $N=$4 extremal Kerr black hole, it still maintain the extremality by zero radial momentum particle absorption. The Kerr black hole makes its outer horizon smaller in the very limited particle radial momentum, and it is almost impossible when the black hole goes to a extremal case. The making smaller horizon is difficult, even if it is possible. Through non-zero radial momentum particle, it is broken to two horizon cases. Therefore, the extremal Kerr black hole cannot not be observed in the real spacetime. Note that the geodesic particle absorption in $N=$4 case can change the black hole properties in the range of $\delta M_B \geq \delta a$\,. For $N=$5 MP black holes, the horizon size changes are dependent on a spin value and particle radial momentums. If the black hole spin is in $0<a<\sqrt{M}$\,, the horizon size becomes smaller for any particle momentum for the same rotating cases and larger for $p^r<\frac{2r_h^2 aL}{\Sigma(a^2-r_h^2)}$ for the opposite rotating cases. If the black hole spin is in $\sqrt{M}<a<\sqrt{2M}$\,, the horizon becomes larger than before $p^r >\frac{2r_h^2 aL}{\Sigma(a^2-r_h^2)}$ for the same rotating cases and any values of momenta for the opposite rotating cases. Simply, the reaction of outer horizon change for particle momenta is divided into $0<a<\sqrt{M}$ and $\sqrt{M}<a<\sqrt{2M}\,$. A particle can make smaller horizon in $0<a<\sqrt{M}$, but it is harder and harder in $\sqrt{M}<a<\sqrt{2M}\,$. Finally, the smaller horizon is impossible when the spin goes to $a\sim \sqrt{2M}$ range. Therefore, the naked ring singularity formed at $a=\sqrt{2M}$ can be avoided by the controlling particle momentum and does not exist in real universe. For $N\geq 6\,$, these have no restriction on a spin value, so we increase a spin value to very large. These black hole becomes ultra spinning cases called black membranes when $a\gg r_h$\,. To achieve ultra-spinning case, the horizon becomes smaller and the spin parameter becomes bigger value after particle absorption. The condition for making smaller horizon satisfies both of them. For the same rotating direction, any radial and angular momenta particle makes the horizon size smaller after the absorption. The range of a radial momentum satisfied this condition becomes tighter for a larger spin value. So, making black membranes from MP black holes is looked as not favorable cases, but it is possible, ideally. For the opposite rotating cases, the horizon size becomes smaller when $|p^r|>\frac{(N-3)r_h^2+(N-5)a^2}{(4-N)r_h^2+(6-N)a^2}aL$\,. Therefore, $N\geq 6$ MP black holes can be achieved to ultra spinning by absorbing the specific particle. For 5-dimensional two-spin MP black hole, through a reversible process, the MP black holes at the spin upper bounds still stay at the spin upper bounds for the changed mass. However, through an irreversible process, the MP black holes at the spin upper bounds are broken to two horizon cases. For over 6 dimensions and two or multi spins, the finding of the spin upper bounds is a difficult problem, because it is dependent on the dimensions and number of spins. We expect that the geodesic particle absorptions still obey our results.

The meaning of the entropy only changed by a particle radial momentum is originated from the microstate counting of the black hole. The Kerr/CFT correspondence is one of applications can explain this situation. The correspondence working on a extremal Kerr black hole is applied to reversible process in a particle absorption to maintain the extremality. It only changes black hole rotation energy which produces the change of black hole mass, not the entropy. Therefore, the change of microstates of the black hole with invariant degeneracy is shown by this process. To avoid confusion, we rewrite the correspondence by the irreducible mass. As a result, the process changes right-hand side momentum and temperature, that is not physically favorable. From these changes, the vacuum and temperature of the extremal Kerr black hole is still not changed. Starting this conclusion, we expect that the irreversible process will affect to its vacuum and temperature to change the entropy, and it will be written as particle momenta. Our work may contribute to understanding the entropy phase diagram of black ring and black Saturn, Penrose process in higher dimensions, as well as the Kerr/CFT conjecture applications after absorption.\\

{\bf Acknowledgments}

This work was supported by the National Research Foundation of Korea(NRF) grant funded by the Korea government(MEST) through the Center for Quantum Spacetime(CQUeST) of Sogang University with grant number 2005-0049409.

\end{document}